\documentclass[12pt]{article}
\usepackage{amsmath,amsfonts,amssymb,latexsym}

\numberwithin{equation}{section}
\def\be{\begin{equation}}
\def\ee{\end{equation}}
\def\bq{\begin{eqnarray}}
\def\eq{\end{eqnarray}}
\def\beq{\begin{eqnarray*}}
\def\eeq{\end{eqnarray*}}

\def\r{\rho}
\def\a{\alpha}
\def\b{\beta}
\def\g{\gamma}
\def\G{\Gamma}
\def\d{\delta}
\def\k{\kappa}

\def\l{\lambda}

\def\m{\mu}

\def\pa{\partial}
\def\r{\rho}
\def\z{\zeta}

\def\ep{\epsilon}
\newcommand{\GA}{\alpha}
\newcommand{\GB}{\beta}
\newcommand{\GG}{\gamma}
\newcommand{\GD}{\delta}
\newcommand{\GE}{\epsilon}

\begin{document}
\title{\huge{The radiation instability in modified gravity}}
\author{\Large{\textsc{Spiros Cotsakis}$^{1,2}$\thanks{skot@aegean.gr}\, and \textsc{Dimitrios Trachilis}$^3$\thanks{Dimitrios.Trachilis@aum.edu.kw}}\\ \\
$^1$Institute of Gravitation and Cosmology\\ RUDN University\\
ul. Miklukho-Maklaya 6, Moscow 117198, Russia\\ \\
$^2$Research Laboratory of Geometry,\\  Dynamical Systems  and Cosmology\\
University of the Aegean\\ Karlovassi 83200, Samos, Greece\\ \\
$^3$College of Engineering and Technology\\
American University of the Middle East, Kuwait}
\date{March 2021}
\maketitle

\newpage
\begin{abstract}
\noindent  We study  the problem of the instability of inhomogeneous radiation universes in quadratic lagrangian theories of gravity written as a system of evolution equations with constraints. We construct formal series expansions and show that the resulting solutions have a smaller number of arbitrary functions than that required in a general solution. These results continue to hold for more general polynomial extensions of general relativity.
\end{abstract}
\newpage
\tableofcontents
\newpage
\section{Introduction}
The Friedman cosmologies are a central theme in modern physical cosmology because they provide an excellent comparison with cosmological observations. The initial singularity in these models is, however, a feature not only generic in general relativity, but also one that persists when these models are analysed in geometric extensions of general relativity such as modified gravity. An important  feature of these generalized models that was  known very early is that if some special FRW solution avoided the initial singularity in the past, it failed to approach the Friedman solution of general relativity and experienced a  curvature singularity in the future (cf. \cite{bo83} and earlier references therein).

A basic such model is the flat, isotropic, radiation universe, which  in the context of extensions of general relativity possesses interesting further properties such as the non-existence of horizons  near the singularity \cite{fhh}, \cite{fb}, and, for certain non-analytic choices of lagrangian, the avoidance of the initial singularity. The existence and (in)stability properties of the flat, radiation solution in analytic  $f(R)$ gravity, on the other hand,  is a well-studied problem (cf. \cite{bo83}, \cite{cf1}, \cite{cf2}, and references therein). It is  known that the singular behaviour of this solution towards the past singularity persists and is stable in the sense that all non-flat, radiation solutions of the same theory asymptotically approach the flat one at early times \cite{ckt}. The radiation stage is also very important in recent studies of dark energy, in particular, of the radiation-matter transition in modified gravity, and a very prolonged radiation era may result in incompatibilities of some modified gravity models with current CMB observations \cite{ame1}-\cite{ame2}.

However, the past-instability conjecture states that all homogeneous and isotropic solutions of general relativity as in general past-unstable when viewed as solutions of higher order gravity \cite{cf2}. Such an instability is a prerequisite in any attempt to examine the cosmological viability of geometric extensions and generalizations of general relativity through arguments relying on a transition between different isotropic solutions. These transitions are  needed in dark energy models based on  relativistic extensions. In this paper, we prove this conjecture for the case of radiation solutions of general relativity in the framework of the quadratic lagrangian theory $R+\ep R^2$. The method of proof consists in constructing  formal series expansions and counting the free functions in the resulting solutions. Then we find that the final solution so constructed cannot have the full number of such functions to qualify as a general solution of the evolution and constraint equations of the problem.

The plan of this paper is as follows. In the next Section, we write down the basic system of evolution and constraint equations for the quadratic $f(R)$ theory as a dynamical system for the two tensors $\gamma, K$, and high-order ones $D,W$ od the space slices, and give a count of the total number of  arbitrary data for the radiation solution to be regarded as a general solution of the theory. In Section 3, we give a new proof of a well known result about the existence of radiation isotropic solutions in quadratic gravity, and provide formal series expansions for the metric and the extrinsic and Ricci curvatures. In Section 4, we calculate the asymptotic structure of the various terms in the basic system of evolution and constraint equations, and in Section 5 we count the free data present in the general perturbation of the radiation solution. Alongside, we prove several technical results that are important for the function counting arguments used in this work, but more importantly because they show that our conclusions  have a wider significance. A discussion of these implications appears in the last Section.
\section{Splittings}
In this Section, we write down the basic dynamical system  which consists of evolution and constraint equations for the quadratic lagrangian theory $f(R)=R+\ep R^2$ written in a splitted form in suitable coordinates as in a Cauchy problem formulation of the theory. We then count the free functions necessary for the radiation solution to be a general solution of the theory, and this number turns out to be 20 such functions.
\subsection{Metric}
We consider a spacetime $(\mathcal{V},g_{ij})$ where $\mathcal{V}=\mathbb{R}\times\mathcal{M}$, with $\mathcal{M}$ being an orientable 3-manifold, and $g_{ij}$ is a Lorentzian metric, analytic and
with signature $(+,-,-,-)$. We also consider a diffeomorphism $\phi:\mathcal{V}\longrightarrow \mathcal{M}\times I, I\subseteq \mathbb{R}$, such that, the submanifolds $\phi^{-1}(\{t\}\times \mathcal{M})=\mathcal{M}_{t}, t\in I$ are spacelike and the curves $\phi^{-1}(\{x\}\times I), x\in \mathcal{M}$ are timelike. Such a frame is called a Cauchy adapted frame \cite{ycb}. A vector field $\partial/\partial t$ on $\mathcal{V}$ is defined by these curves and it can be decomposed into normal and parallel components relative to the slicing as follows:
\be
\partial_t=Nn+\overline{N}.
\label{eq:lapse_shift}
\ee
Here $N$ is a positive function on $\mathcal{V}$, $n$ is a future-directed, unit, normal field, and $\overline{N}$ is a vector field tangent to the slices $\mathcal{M}_{t}$. As usual we call $N$ the lapse function and $\overline{N}$ the shift vector field. Then we set,
\be
\overline{N}=N^\a \partial_\a,
\label{eq:shift vector field}
\ee
and due to the fact that $\partial_t$ is a timelike vector field on $\mathcal{V}$ we obtain,
\bq
g_{00}&=&g(\partial_t,\partial_t)=N^2 + N^\a N_\a>0,
\label{eq:positive_g_00}\\
g_{0 \a}&=&g(\partial_t, \partial_\a)=N_\a, \quad g^{0 \a}=N^\a/N^2,
\label{eq:shift}
\eq
so that,
\be
ds^2=N^{2}dt^{2} - (\g_{\a\b}-\frac{N_\a N_\b}{N^2 + N^\g N_\g})(dx^\a + N^\a dt)(dx^\b + N^\b dt).
\label{eq:ghj}
\ee
Using the choice (geodesic slicing)  $N=1, N^\a=0$, that is $g_{00}=1,g_{0\a}=0$, we arrive at  the synchronous system of local coordinates  (the spacetime $\mathcal{V}$ is a sliced spacetime in this system) where
 the {4}-metric $g_{ij}$ is
\be
ds^2=dt^2-\g_{\a\b}dx^\a dx^\b,
\label{eq:Synchronous_ds^2}
\ee
with the time $t$ measuring proper time.

\subsection{Radiation fluid}
We assume a radiation content for the model treated below. In particular, we define a cosmological perfect fluid is defined to be a continuous distribution of matter with energy-momentum tensor of the form,
\be
T^i_j=(\rho+p)u^i u_j-p\d^i_j,
\label{eq:general perfect fluid}
\ee
where $\rho$ is the energy density and $p$ is the pressure satisfying the equation of state,
\be
p=w\rho,
\label{eq:general equation of state}
\ee
where $w\in \mathbb{R}$ is the equation of state parameter. In this paper, we take $w=1/3$, although many of the results hold obviously true for $w$ in an interval. We are not going to repeat this statement every time is true. We assume that the velocity  vector,
\be
u^i=\left(\dfrac{dt}{ds},\dfrac{dx^\a}{ds}\right)=(u^0,u^\a),
\label{eq:velocity}
\ee
associated with  any timelike curve parametrized by the proper time $s$ has  unit length,
\be
1=u_i u^i= u_0 u^0 + u_\a u^\a.
\label{eq:velocities identity}
\ee
The $4$-velocity $u^i$ is used for a covariant description of continuous matter distributions given by the energy-momentum tensor $T_{ij}$.

\subsection{Evolution and constraint equations}
We start from the field equations of the higher-order gravity theory derived from an analytic lagrangian $f(R)$,
\be
L_{ij}=f'(R)R_{ij} -\frac{1}{2}f(R)g_{ij} -\nabla_i\nabla_j f'(R) +g_{ij}\Box_g f'(R)=8\pi G T_{ij},
\label{eq:FEs_rad}
\ee
where $G$ is Newton's gravitational constant, and $T_{ij}$ is the energy-momentum tensor satisfying the conservations laws of energy and momentum, given by the equations of motion,
\be
\nabla_i T^i_j=0.
\label{eq:conservation laws}
\ee
In this paper, we specialize in the quadratic theory $f(R)=R+\ep R^2$, where the field equations (\ref{eq:FEs_rad}) in a Cauchy adapted frame split as follows:
\be
L_{00} =(1+ 2\ep R)R_{00} -\frac{1}{2}(1+\ep R)R +2\ep g^{\a\b} \nabla_\a \nabla_\b R =8\pi G T_{00},
\label{eq:Loo}
\ee
\be
L_{0\a} =(1 +2\ep R)R_{0\a} -2\ep \nabla_0 \nabla_\a R =8\pi G T_{0 \a},
\label{eq:Loa}
\ee
\be
L_{\a\b} = (1 +2\ep R)R_{\a\b} -\frac{1}{2}(1+\ep R)R g_{\a\b} -2\ep \nabla_\a \nabla_\b R +2\ep g_{\a\b} \Box_g R =
8\pi G T_{\a\b}.
\label{eq:Lab}
\ee
Here, using standard methods for splitting the Riemann tensor and the connection \cite{ycb}, the various components of the Ricci $4$-curvanture are given by the formulae,
\be
R_{00} = -\frac{1}{2}\partial_t K  -\frac{1} {4}K_\a^\b K_\b^\a,
\label{eq:R00}
\ee
\be
R_{0\a}  = \frac{1}{2} (\nabla_\b K^\b_\a - \nabla_\a K),
\label{eq:R0a}
\ee
and also
\be
R_{\a\b}= P_{\a\b} +\frac{1} {2} \partial_t K_{\a\b} +\frac{1} {4}K K_{\a\b} -\frac{1}{2}K^\g_\a K_{\b\g}.
\label{eq:Rab}
\ee
Here, $P_{\a\b}$ denotes the three-dimensional Ricci tensor associated with $\g_{\a\b}$, and $K=\textrm{tr}_{\gamma}K_{\a\b}=\gamma^{\a\b}K_{\a\b}$ is the mean curvature of the slices $\mathcal{M}_{t}$, with $K_{\a\b}$ being the extrinsic curvature (or, second fundamental form) of the spatial slices $\mathcal{M}_{t}$,  defined by the \emph{first variational equation}
\be
\partial_t \g_{\a\b} = K_{\a\b}.
\label{eq:pg}
\ee
Because of the higher-than-second-order derivatives present in the field equations, we need to introduce here further important symbols, namely,  the symmetric \emph{acceleration tensor}  $D_{\a\b}$  through the corresponding \emph{second variational equation}
\be
\partial_t K_{\a\b} = D_{\a\b},
\label{eq:pK}
\ee
as well as the \emph{jerk tensor} $W_{\a\b}$ satisfying the  equation,
\be
\partial_t D_{\a\b} = W_{\a\b}.
\label{eq:pD}
\ee
Then the final evolution evolution equation of the quadratic theory, called here the \emph{snap} equation, becomes,
\begin{eqnarray}
&&\partial_t W =\frac{1}{6\GE}(8\pi G T^\a_\a + \frac{1}{2} P +\frac{1}{8} K^2  -\frac{5}{8} K^{\GA\GB} K_{\GA\GB} +D ) + \nonumber \\
&&\frac{1}{6}[P^2 +\frac{1}{4}P K^2 -\frac{1}{4}P K^{\GA\GB} K_{\GA\GB} +\frac{1}{32} K^4 -\frac{1}{16}K^2 K^{\GA\GB} K_{\GA\GB}  -\nonumber \\
&&6K K^\GA_\GB K^\GB_\GG K^\GG_\GA -\frac{99}{32} (K^{\GA\GB} K_{\GA\GB})^2 +27 K^\GA_\GB K^\GB_\GG K^\GG_\GD K^\GD_\GA +9 K K^{\GA\GB} D_{\GA\GB}  -  \nonumber \\
&&57 K^\GA_\GB K^\GB_\GG D^\GG_\GA  +\frac{13}{2}D K^{\GA\GB} K_{\GA\GB} -\frac{7}{2}D^2 +15 D^{\GA\GB} D_{\GA\GB}  -3K W   +\nonumber \\
&&15K^{\GA\GB} W_{\GA\GB}  -6\partial_t (\partial_t P) -\nonumber \\
&&4\g^{\GA\GB} \nabla_\GA \nabla_\GB (-P -D +\frac{3}{4} K^{\GA\GB} K_{\GA\GB} -\frac{1}{4}K^2)].
\label{eq:pW}
\end{eqnarray}
In addition, we find that the theory contains \emph{constraint equations}, which  read as follows,

\emph{Hamiltonian Constraint}
\begin{eqnarray}
\mathcal{C}_0 :&&\frac{1}{2}P  +\frac{1}{8}K^2 -\frac{1}{8}K^{\a\b} K_{\a\b} + \nonumber \\
&&\ep [-\frac{1}{2}P^2  -\frac{1}{4}P K^2 +\frac{1}{4}P K^{\a\b} K_{\a\b}  -\frac{1}{32}K^4 +\frac{1}{16}K^2 K^{\a\b} K_{\a\b} + \nonumber\\
&&\frac{3}{32}(K^{\a\b} K_{\a\b})^2 -\frac{1}{2}D K^{\a\b} K_{\a\b} +\frac{1}{2}D^2     - \nonumber\\
&&2\g^{\a\b} \nabla_\a \nabla_\b(-P - \frac{1}{4}K^2 +\frac{3}{4}K^{\g\d} K_{\g\d}  -D)] =8\pi G T_{00},
\label{eq:hamiltonian}
\end{eqnarray}

\emph{Momentum Constraint}
\begin{eqnarray}
\mathcal{C}_\a :&&\frac{1}{2} (\nabla_\b K^\b_\a - \nabla_\a K) + \nonumber \\
&&\ep [(-P  -\frac{1}{4}K^2 +\frac{3}{4}K^\d_\g K^\g_\d -D) (\nabla_\b K^\b_\a - \nabla_\a K)  - \nonumber \\
&&\nabla_\a (-2\partial_t P + K K^{\g\d} K_{\g\d} -3 K^\b_\g K^\g_\d K^\d_\b -K D +5K^{\g\d} D_{\g\d} -2W)]\nonumber \\&&=
8\pi G T_{0\a}.
\label{eq:momentum}
\end{eqnarray}
We note that in the two constraints above the brackets multiplied by $\epsilon$ contain the extra higher-order terms not present in the Einstein equations.
\subsection{Counting}
The four evolution equations (\ref{eq:pg}), (\ref{eq:pK}), (\ref{eq:pD}) and (\ref{eq:pW})  together with the constraints  (\ref{eq:hamiltonian}) and (\ref{eq:momentum}), the equation of state (\ref{eq:general equation of state}), and the identity (\ref{eq:velocities identity}), constitute the basic system studied in this paper. They describe the
time development $(\mathcal{V},g_{ij})$ of any initial data set $(\mathcal{M}_{t},\g_{\a\b},K_{\a\b},D_{\a\b},W_{\a\b})$ together with the quantities $p,\rho,u^i$ in the present theory. The vacuum parts of the equations are identical to those found in \cite{ct16}, but we give them here for easy reference. This system satisfies the Cauchy-Kovalewski property, but since the proof is identical to that in Ref. \cite{ct16}, we refer the reader to that reference.

Based on the previous relations, we end up with a dynamical system consisting of $30$ arbitrary functions, the $24$ initial data $(\g_{\a\b},K_{\a\b},D_{\a\b},W_{\a\b})$ together with the $6$ functions $p,\rho,u^i$ satisfying evolution equations (\ref{eq:pg}), (\ref{eq:pK}), (\ref{eq:pD}) and (\ref{eq:pW}) as well as the four constraint equations (\ref{eq:hamiltonian}) and (\ref{eq:momentum}) on each slice $\mathcal{M}_{t}$. The quantities $p,\rho$ satisfy the equation of state (\ref{eq:general equation of state}), and the velocities the identity (\ref{eq:velocities identity}).

Hence, the number of arbitrary functions that have to be specified initially is equal to $30-4-1-1-4=20$, that is  from the initial $30$ functions we have to subtract in turn $4$ from the constraint equations, $1$ from the equation of state, $1$ from the identity for the velocities, and the  $4$ diffeomorphisms. This number 20 is the correct number for any kind of perfect fluid matter including radiation ($w=1/3$), also consistent with that calculated in \cite{ba} (corresponding to the notation $F=1,S=0$ of that paper).

\section{Curvatures}
In this Section, after giving a new proof of the existence of  exact radiation solutions in the quadratic theory, we calculate the necessary formal expansions for the mean and extrinsic curvatures, acceleration and jerk tensors, the spatial Ricci curvature, and so for the Ricci 4-curvature components and the scalar curvature. These expansions will be used in the next Section when we expand the constraint and evolution equations.

\subsection{The exact radiation solution}
For the FRW metric,
\be
{ds}^2 = {dt}^2 - {a^2}(t)\left[ \dfrac{{dr}^2}{1-kr^2} + r^2({d\theta}^2 + {\sin}^2 \theta {d\phi}^2) \right],
\label{rwmetric}
\ee
the off-diagonal components of the Ricci tensor are equal to zero, and the diagonal components are given by,
\bq
R_{00} &=& -3\dfrac{\ddot{a}}{a},\\
\label{R00sf}
R_{11} &=& \dfrac{a\ddot{a}+2{\dot{a}}^2 +2k}{1-kr^2},\\
\label{R11sf}
R_{22} &=& \dfrac{a\ddot{a}+2{\dot{a}}^2 +2k}{r^2},\\
\label{R22sf}
R_{33} &=& \dfrac{a\ddot{a}+2{\dot{a}}^2 +2k}{r^2 {\sin}^2 \theta},
\label{R33sf}
\eq
while for the scalar curvature we obtain,
\be
R = -6\left( \dfrac{\ddot{a}}{a} + \dfrac{{\dot{a}}^2}{a^2} + \dfrac{k}{a^2} \right),
\label{Rsf}
\ee
where dot denotes differentiation with respect to $t$. Working  with the lagrangian $f(R)=R + \ep R^2$ and the field equations (\ref{eq:FEs_rad}), in additon to  the Eqns. (\ref{R00sf})-(\ref{Rsf}), we find that,
\bq
\pa_t R &=& -6\left( \dfrac{\dddot{a}}{a} + \dfrac{\dot{a}\ddot{a}}{a^2} - 2\dfrac{{\dot{a}}^3}{a^3} - 2k\dfrac{\dot{a}}{a^3} \right), \\
\label{pa_t Rsf}
{\pa_t}^2 R &=& -6\left( \dfrac{a^{(4)}}{a} + \dfrac{{\ddot{a}}^2}{a^2} - 8\dfrac{{\dot{a}}^2 \ddot{a}}{a^3} + 6\dfrac{{\dot{a}}^4}{a^4} - 2k\dfrac{\ddot{a}}{a^3} + 6k\dfrac{{\dot{a}}^2}{a^4} \right),
\label{pa_t2 Rsf}
\eq
as well as,
\bq
\nabla_1 \nabla_1 R &=& \dfrac{6}{1-kr^2}\left( \dot{a} \dddot{a} + \dfrac{{\dot{a}}^2 \ddot{a}}{a} - 2\dfrac{{\dot{a}}^4}{a^2} - 2k\dfrac{{\dot{a}}^2}{a^2} \right), \\
\label{nabla1nabla1Rsf}
\nabla_2 \nabla_2 R &=& 6r^2 \left( \dot{a} \dddot{a} + \dfrac{{\dot{a}}^2 \ddot{a}}{a} - 2\dfrac{{\dot{a}}^4}{a^2} - 2k\dfrac{{\dot{a}}^2}{a^2} \right), \\
\label{nabla2nabla2Rsf}
\nabla_3 \nabla_3 R &=& 6r^2\sin{\theta} \left( \dot{a} \dddot{a} + \dfrac{{\dot{a}}^2 \ddot{a}}{a} - 2\dfrac{{\dot{a}}^4}{a^2} - 2k\dfrac{{\dot{a}}^2}{a^2} \right), \\
\label{nabla3nabla3Rsf}
\nabla_\a \nabla_\b R &=& 0, \quad \textrm{if} \quad \a \neq \b.
\label{nablaanablabRsf}
\eq
Then, the $00$-component and $\a \a$-components of  the generalized Friedmann equations for the energy density $\rho$ and the pressure $p$, are given by \cite{ckt},
\be
\dfrac{8\pi G\rho}{3}=\dfrac{k+\dot{a}^2}{a^2} +6\ep \left( 2\dfrac{\dot{a}\dddot{a}}{a^2} +2\dfrac{\dot{a}^2 \ddot{a}}{a^3} -\dfrac{\ddot{a}^2}{a^2} -3\dfrac{\dot{a}^4}{a^4} -2k\dfrac{\dot{a}^2}{a^4} +\dfrac{k^2}{a^4} \right),
\label{gen Friedmann eq density}
\ee
and,
\bq
8\pi G p&=&-2\dfrac{\ddot{a}}{a}- \dfrac{\dot{a}^2 +k}{a^2} \nonumber\\
&+&6\ep \left( -2\dfrac{a^{(4)}}{a} -4\dfrac{\dot{a} \dddot{a}}{a^2} -3\dfrac{\ddot{a}^2}{a^2} +12\dfrac{\dot{a}^2 \ddot{a}}{a^3} -3\dfrac{\dot{a}^4}{a^4} +4k\dfrac{\ddot{a}}{a^3} -2k\dfrac{\dot{a}^2}{a^4} +\dfrac{k^2}{a^4} \right),
\label{gen Friedmann eq pressure}
\eq
where the general relativity case is recovered when $\ep= 0$. The parts not multiplied by the coefficient $6\ep$ in the right-hand-sides are identical to the standard general relativistic expressions $\dfrac{8\pi G\rho_{GR}}{3}$ and $8\pi G p_{GR}$ respectively. The extra parts multiplied by $6\ep$ become zero if we substitute the standard forms for radiation, namely, for $k=-1,0,+1$ are respectively (cf. \cite{on}, chap. 12):
\be
a=\sqrt{c}[(1+\dfrac{t}{\sqrt{c}})^2 -1]^{1/2}=(2\sqrt{c}t+t^2)^{1/2},
\label{sol rad GR k=-1}
\ee
\be
a=(4c)^{1/4}t^{1/2},
\label{sol rad GR k=0}
\ee
\be
a=\sqrt{c}[1-(1-\dfrac{t}{\sqrt{c}})^2]^{1/2}=(2\sqrt{c}t-t^2)^{1/2},
\label{sol rad GR k=+1}
\ee
where,
\be
c=\dfrac{8\pi G \rho a^4}{3}
\label{eq c}
\ee
and,
\be
\rho a^4=\textrm{constant},
\label{eq ra^4}
\ee
which expresses the conservation of mass-energy (\ref{eq:conservation laws}). Therefore  the exact radiation solutions of the Friedmann equations for all $k$  are also exact solutions of the generalized Friedmann equations. This is a different, direct, proof of a  well known result about homogeneous and isotropic solutions of general relativity being also solutions of an $f(R)$-theory with $f(0)=0\neq f'(0)$ and with  trace-free matter (cf. \cite{bo83}, Section 4.1).

\subsection{Radiation metric expansions}
Since in the quadratic theory  the asymptotic form of the scale factor for the exact radiation solution of the previous Subsection is given by,
\be
a\propto t^{1/2},
\label{a prop t^1/2}
\ee
we assume a formal series representation of the spatial metric for a generic solution in this case of the form,
\bq
\g_{\a\b} &=& \g^{(1)}_{\a\b}t + \g^{(2)}_{\a\b}t^2 + \g^{(3)}_{\a\b}t^3 + \g^{(4)}_{\a\b}t^4 + \cdots \nonumber \\
          &=& a_{\a\b}t + b_{\a\b}t^2 + c_{\a\b}t^3 + d_{\a\b}t^4 + \cdots,
\label{spatial metric rad}
\eq
where the $\g^{(1)}_{\a\b}=a_{\a\b} , \g^{(2)}_{\a\b}=b_{\a\b} , \g^{(3)}_{\a\b}=c_{\a\b} , \g^{(4)}_{\a\b}=d_{\a\b},\cdots$ are arbitrary, nontrivial analytic functions of the space coordinates.

It is not difficult to count the number of arbitrary functions present in this expansion. In Section \ref{gthan4}, we shall show that keeping all terms of order higher than four leads to qualitatively the same results, and therefore we  can  keep only  terms of order up to four in this series and remove the dots at the end from now on. Hence, before substitution to the evolution and constraint higher order equations, the basic metric expansion   (\ref{spatial metric rad}) contains  $24$ degrees of freedom ($6$ arbitrary  functions in each spatial matrix $\g^{(n)}_{\a\b},n=1,\cdots,4$).

To calculate the formal expansion of the inverse metric $\g^{\a\b} = \sum_{n=-1}^{\infty}(\g^{\a\b})^{(n)} t^n$, we use the identity $\g_{\a\b}\g^{\b\g}=\d^\g_\a$. Then  for the components $ \g^{(\mu)\,\a\b},\mu=1,\cdots,4,$  we obtain,
\be
\g^{\a\b} = \dfrac{1}{t}a^{\a\b} - b^{\a\b} + t(-c^{\a\b} + b^{\a\g}b^\b_\g) +
t^2(-d^{\a\b} + b^{\a\g}c^\b_\g - b^{\a\g}b^\d_\g b^\b_\d + c^{\a\g}b^\b_\g).
\label{eq:3diminvmetric_rad}
\ee
Here $a_{\a\b}a^{\b\g}=\d_\a^\g$ and the indices in $b_{\a\b},c_{\a\b},d_{\a\b}$ are raised by $a^{\a\b}$.

Using the metric  expansion  ($\ref{spatial metric rad}$), we may calculate the coefficients of any tensor $X$ in the expansion
\be
X_{\a\b} = \sum X^{(n)}_{\a\b}\;t^n,
\label{X tensor}
\ee
where the values of $n$ depend on the tensor. In the remaining of this Section, we do this for the various curvatures.

\subsection{Extrinsic curvature, acceleration and jerk tensors}
For the extrinsic curvature, in terms of the data $a_{\a\b},b_{\a\b},c_{\a\b},d_{\a\b}$, we have explicitly,
\be
K_{\a\b} = \partial_t \g_{\a\b}= a_{\a\b} + 2tb_{\a\b} + 3t^2c_{\a\b} + 4t^3d_{\a\b},
\label{eq:Kab_rad}
\ee
and so for the mixed components we find,
\be
K_\b^\a = \dfrac{1}{t}\d^\b_\a + b^\b_\a + t(2c^\b_\a - b^\b_\g b^\g_\a) +
t^2(3d^\b_\a - 2b^\b_\g c^\g_\a - c^\b_\g b^\g_\a + b^\b_\g b^\g_\d b^\d_\a).
\label{eq:Kmixed_rad}
\ee
Additionally, since the spatial metric determinant satisfies $\g > 0$ \cite{ll}, for
the mean curvature,
\be
K=K_\a^\a=\g^{\a\b} \partial_t\g_{\a\b}=\partial_t  ln(\g),
\label{mean curvature_rad}
\ee
we have the expansion,
\be
K = \dfrac{3}{t} + b + t(2c - b^\b_\a b^\a_\b) + t^2(3d - 3b^\b_\a c^\a_\b + b^\b_\g b^\g_\a b^\a_\b).
\label{eq:trK_rad}
\ee
The corresponding  expressions for the coefficients $(K^{\a\b})^{(n)}$ of the fully contravariant symbols are,
\be
K^{\a\b} = \dfrac{1}{t^2}a^{\a\b} + (c^{\a\b} - b^{\b\g}b^\a_\g) +
t(2d^{\a\b} - 2b^\b_\g c^{\g\a} - 2b^{\a\g}c^\b_\g + 2b^{\a\g}b^\d_\g b^\b_\d).
\label{eq:K^ab_rad}
\ee
We  can now find the various components of the acceleration and jerk tensors to the required order. For the acceleration we have,
\be
D_{\a\b} = \pa_t K_{\a\b} = 2b_{\a\b} + 6c_{\a\b}t + 12d_{\a\b}t^2,
\label{eq:Dab_rad}
\ee
and for the mixed components we find,
\be
D^\b_\a = \frac{2}{t}b^\b_\a + 2(3c^\b_\a -b^\b_\g b^\g_\a) +
2t(6d^\b_\a -3b^\b_\g c^\g_\a -c^\b_\g b^\g_\a
+b^\b_\d b^\d_\g b^\g_\a),
\label{eq:Dmixed_rad}
\ee
so that,
\be
D = \frac{2}{t}b + 2(3c -b^\b_\a b^\a_\b) + 2t(6d -4b^\b_\a c^\a_\b + b^\b_\g b^\g_\a b^\a_\b),
\label{eq:trD_rad}
\ee
where the trace is given by,
\be
D=D^\a_\a=\g^{\a\b} \pa_t K_{\a\b}.
\label{eq:trace_D_rad}
\ee
The fully contravariant components are given by,
\be
D^{\a\b} = \frac{2}{t^2}b^{\a\b} + \frac{2}{t}(3c^{\a\b} - 2b^\b_\g b^{\a\g}) +
2(6d^{\a\b} - 4b^\b_\g c^{\a\g} - 4b^\a_\g c^{\g\b} + 3b^{\a\g} b^\d_\g b^\b_\d).
\label{eq:D^ab_rad}
\ee
For the jerk, we have
\be
W_{\a\b} = \pa_t D_{\a\b} = 6c_{\a\b} + 24d_{\a\b}t,
\label{eq:Wab}
\ee
so that,
\be
W = \dfrac{6}{t}c + 6(4d - b^\b_\a c^\a_\b),
\label{eq:trW_rad}
\ee
with
\be
W=W^\a_\a=\g^{\a\b} \pa_t D_{\a\b}.
\label{eq:trace_W_rad}
\ee
This completes the extrinsic curvature calculations.
\subsection{Ricci and scalar curvatures}
For the three-dimensional Ricci tensor $P_{\a\b}$ and its trace $P=\textrm{tr}_\gamma P_{\a\b}$ we may also apply the above method to get the corresponding expansions.
The tensor $P_{\a\b}$ satisfies,
\be
P_{\a\b} = \pa_\m \G^\m_{\a\b} - \pa_\b \G^\m_{\a\m} + \G^\m_{\a\b}\G^\ep_{\m\ep} - \G^\m_{\a\ep}\G^\ep_{\b\m},
\label{three-dimensional Ricci tensor 2}
\ee
and so finding the expansion of $P_{\a\b}$ depends on the corresponding ones for the Christoffel symbols,
\be
\G^\m_{\a\b} = \dfrac{1}{2}\g^{\m\ep}(\pa_\b \g_{\a\ep} + \pa_\a \g_{\b\ep} - \pa_\ep \g_{\a\b}),
\label{Christoffel symbols}
\ee
The final result which we now prove is as follows,
\be
P_{\a\b} = \widetilde{P}_{\a\b} + tH_{\a\b} + t^2 I_{\a\b} + t^3 J_{\a\b},
\label{finally three-dimensional Ricci tensor}
\ee
where the coefficients $\tilde{P},H,I,J$ are purely spatial functions of the metric coefficients $a,b,c,d$. To show this result,  we use  the metric  series (\ref{spatial metric rad}) in the Christoffel symbols (\ref{Christoffel symbols}) to obtain,
\bq
\G^\m_{\a\b} &=& \dfrac{1}{2}\big[\dfrac{1}{t}a^{\m\ep} - b^{\m\ep} + t(-c^{\m\ep} + b^{\m\g}b^\ep_\g) +
t^2(-d^{\m\ep} + b^{\m\g}c^\ep_\g + c^{\m\g}b^\ep_\g - b^{\m\g}b^\d_\g b^\ep_\d)\big] \nonumber \\
& \times & \left[t(\pa_\b a_{\a\ep} + \pa_\a a_{\b\ep} - \pa_\ep a_{\a\b}) +
t^2(\pa_\b b_{\a\ep} + \pa_\a b_{\b\ep} - \pa_\ep b_{\a\b}) \right. \nonumber \\
&+& \left. t^3(\pa_\b c_{\a\ep} + \pa_\a c_{\b\ep} - \pa_\ep c_{\a\b}) +
t^4(\pa_\b d_{\a\ep} + \pa_\a d_{\b\ep} - \pa_\ep d_{\a\b})\right].
\label{spatial Christoffel symbols}
\eq
In order to simplify the ensuing calculations we set,
\be
A_{\a\b\ep} = \pa_\b a_{\a\ep} + \pa_\a a_{\b\ep} - \pa_\ep a_{\a\b},
\label{Aabe}
\ee
\be
B_{\a\b\ep} = \pa_\b b_{\a\ep} + \pa_\a b_{\b\ep} - \pa_\ep b_{\a\b},
\label{Babe}
\ee
\be
C_{\a\b\ep} = \pa_\b c_{\a\ep} + \pa_\a c_{\b\ep} - \pa_\ep c_{\a\b},
\label{Cabe}
\ee
\be
D_{\a\b\ep} = \pa_\b d_{\a\ep} + \pa_\a d_{\b\ep} - \pa_\ep d_{\a\b}.
\label{Dabe}
\ee
Then Eq.  (\ref{spatial Christoffel symbols}) becomes,
\be
\G^\m_{\a\b} = \dfrac{1}{2}(\dfrac{1}{t}a^{\m\ep} - b^{\m\ep} + t(\g^{\m\ep})^{(1)} + t^2(\g^{\m\ep})^{(2)})
\times(tA_{\a\b\ep} + t^2 B_{\a\b\ep} + t^3 C_{\a\b\ep} + t^4 D_{\a\b\ep}),
\label{spatial Christoffel symbols ABCD}
\ee
and so,
\bq
\G^\m_{\a\b} &=& \dfrac{1}{2}a^{\m\ep}A_{\a\b\ep} + \dfrac{1}{2}(a^{\m\ep}B_{\a\b\ep} - b^{\m\ep}A_{\a\b\ep})t
+ \dfrac{1}{2}(a^{\m\ep}C_{\a\b\ep} - b^{\m\ep}B_{\a\b\ep} + \g^{(1)\m\ep}A_{\a\b\ep})t^2 \nonumber \\
&+& \dfrac{1}{2}(a^{\m\ep}D_{\a\b\ep} - b^{\m\ep}C_{\a\b\ep} + (\g^{\m\ep})^{(1)}B_{\a\b\ep}
+ (\g^{\m\ep})^{(2)}A_{\a\b\ep})t^3.
\label{spatial Christoffel symbols ABCD series}
\eq
We further set,
\be
\widetilde{\G}^\m_{\a\b} = \frac{1}{2}a^{\m\ep} A_{\a\b\ep},
\label{eq:Gtildemab}
\ee
\be
E^\m_{\a\b} = \frac{1}{2}(a^{\m\ep} B_{\a\b\ep} - b^{\m\ep} A_{\a\b\ep}),
\label{eq:Emab}
\ee
\be
F^\m_{\a\b} = \frac{1}{2}(a^{\m\ep} C_{\a\b\ep} - b^{\m\ep} B_{\a\b\ep} + e^{\m\ep} A_{\a\b\ep}),
\label{eq:Fmab}
\ee
\be
G^\m_{\a\b} = \dfrac{1}{2}(a^{\m\ep}D_{\a\b\ep} - b^{\m\ep}C_{\a\b\ep} + (\g^{\m\ep})^{(1)}B_{\a\b\ep}
+ (\g^{\m\ep})^{(2)}A_{\a\b\ep}),
\label{eq:Gmab}
\ee
so that for the Christoffel symbols we find,
\be
\G^\m_{\a\b} = \widetilde{\G}^\m_{\a\b} + tE^\m_{\a\b} + t^2 F^\m_{\a\b} + t^3 G^\m_{\a\b}.
\label{finally Christoffel symbols}
\ee
Moreover, from (\ref{three-dimensional Ricci tensor 2}) using (\ref{finally Christoffel symbols}) we find,
\bq
P_{\a\b} &=& (\pa_\m \widetilde{\G}^\m_{\a\b} + t \pa_\m E^\m_{\a\b} + t^2 \pa_\m F^\m_{\a\b} + t^3 \pa_\m G^\m_{\a\b})\nonumber\\
&-& (\pa_\b \widetilde{\G}^\m_{\a\m} + t \pa_\b E^\m_{\a\m} + t^2 \pa_\b F^\m_{\a\m} + t^3 \pa_\b G^\m_{\a\m}) \nonumber \\
&+& (\widetilde{\G}^\m_{\a\b} + tE^\m_{\a\b} + t^2 F^\m_{\a\b} + t^3 G^\m_{\a\b})
\times (\widetilde{\G}^\ep_{\m\ep} + tE^\ep_{\m\ep} + t^2 F^\ep_{\m\ep} + t^3 G^\ep_{\m\ep}) \nonumber \\
&-& (\widetilde{\G}^\m_{\a\ep} + tE^\m_{\a\ep} + t^2 F^\m_{\a\ep} + t^3 G^\m_{\a\ep})
\times (\widetilde{\G}^\ep_{\b\m} + tE^\ep_{\b\m} + t^2 F^\ep_{\b\m} + t^3 G^\ep_{\b\m}),
\label{three-dimensional Ricci tensor GEFG}
\eq
and setting further,
\be
\widetilde{P}_{\a_\b} = \pa_\m \widetilde{\G}^\m_{\a\b} - \pa_\b \widetilde{\G}^\m_{\a\m}
+ \widetilde{\G}^\m_{\a\b}\widetilde{\G}^\ep_{\m\ep} - \widetilde{\G}^\m_{\a\ep}\widetilde{\G}^\ep_{\b\m},
\label{eq:Ptildeab}
\ee
\be
H_{\a\b} = \pa_\m E^\m_{\a\b} - \pa_\b E^\m_{\a\m} + \widetilde{\G}^\m_{\a\b}E^\ep_{\m\ep}
+ \widetilde{\G}^\ep_{\m\ep}E^\m_{\a\b} -  \widetilde{\G}^\m_{\a\ep}E^\ep_{\b\m} - \widetilde{\G}^\ep_{\b\m}E^\m_{\a\ep},
\label{eq:Hab}
\ee
\be
I_{\a\b} = \pa_\m F^\m_{\a\b} - \pa_\b F^\m_{\a\m} + \widetilde{\G}^\m_{\a\b}F^\ep_{\m\ep}
+ \widetilde{\G}^\ep_{\m\ep}F^\m_{\a\b} - \widetilde{\G}^\m_{\a\ep}F^\ep_{\b\m} - \widetilde{\G}^\ep_{\b\m}F^\m_{\a\ep}
+ E^\m_{\a\b}E^\ep_{\m\ep} - E^\m_{\a\ep}E^\ep_{\b\m},
\label{eq:Iab}
\ee
\bq
J_{\a\b} &=& \pa_\m G^\m_{\a\b} - \pa_\b G^\m_{\a\m} +\widetilde{\G}^\m_{\a\b}G^\ep_{\m\ep}
+ \widetilde{\G}^\ep_{\m\ep}G^\m_{\a\b} - \widetilde{\G}^\m_{\a\ep}G^\ep_{\b\m} - \widetilde{\G}^\ep_{\b\m}G^\m_{\a\ep}
\nonumber \\
&+& E^\m_{\a\b}F^\ep_{\m\ep} - E^\m_{\a\ep}F^\ep_{\b\m} + F^\m_{\a\b}E^\ep_{\m\ep} - F^\m_{\a\ep}E^\ep_{\b\m},
\label{eq:Jab}
\eq
we obtain the announced  result (\ref{finally three-dimensional Ricci tensor}).

For the mixed three-dimensional Ricci tensor we have,
\be
P^\b_\a = \g^{\b\m}P_{\m\a} = (\dfrac{1}{t}a^{\b\m} - b^{\b\m} + t\g^{(1)\b\m} + t^2 \g^{(2)\b\m})
\times (\widetilde{P}_{\m_\a} + tH_{\m\a} + t^2 I_{\m\a} + t^3 J_{\m\a}),
\label{mixed Pab}
\ee
and setting,
\be
\k^\b_\a = a^{\b\m}H_{\m\a} - b^{\b\m}\widetilde{P}_{\m\a},
\label{eq:kab}
\ee
\be
\l^\b_\a = a^{\b\m}I_{\m\a} - b^{\b\m}H_{\m\a} + (\g^{\b\m})^{(1)}\widetilde{P}_{\m\a},
\label{eq:greekLab}
\ee
\be
\m^\b_\a = a^{\b\m}J_{\m\a} - b^{\b\m}I_{\m\a} + (\g^{\b\m})^{(1)}H_{\m\a} + (\g^{\b\m})^{(2)} \widetilde{P}_{\m\a},
\label{eq:Mab}
\ee
we find that the form of the mixed three-dimensional Ricci tensor is,
\be
P^\b_\a = \dfrac{1}{t}\widetilde{P}^\b_\a + \k^\b_\a + t\l^\b_\a + t^2 \m^\b_\a.
\label{final mixed Pab}
\ee
Finally, we obtain,
\be
P = P^\a_\a = \dfrac{1}{t}\widetilde{P} + \k + t\l + t^2 \m,
\label{final P}
\ee
where obviously,
\be
\k = \d^\a_\b \k^\b_\a,
\label{eq:k}
\ee
\be
\l = \d^\a_\b \l^\b_\a,
\label{eq:greekL}
\ee
\be
\m = \d^\a_\b \m^\b_\a.
\label{eq:M}
\ee

Using the above results, we can express the  components of the Ricci curvature as well as  its trace in terms of the asymptotic data $a_{\a\b},b_{\a\b},c_{\a\b},d_{\a\b}$. Based on Eqns. (\ref{eq:R00}), (\ref{eq:R0a}) and (\ref{eq:Rab}) and our previous results concerning the quantities $P_{\a\b}$, the  components of the Ricci curvature become,
\be
R^0_0 = \frac{3}{4t^2} - \frac{1}{2t}b + (-2c +\frac{3}{4}b^\b_\a b^\a_\b) +
(-\frac{9}{2}d + \frac{7}{2}b^\b_\a c^\a_\b - b^\b_\g b^\g_\a b^\a_\b)t,
\label{eq:R00mixed_rad}
\ee
\bq
R^0_\a &=& \frac{1}{2}(\nabla_\b b^\b_\a - \nabla_\a b) + [(\nabla_\b c^\b_\a -\nabla_\a c)
-\frac{1}{2}\nabla_\b (b^\b_\g b^\g_\a) + \frac{1}{2}\nabla_\a (b^\b_\g b^\g_\b)]t \nonumber \\
&+& [\frac{3}{2}(\nabla_\b d^\b_\a - \nabla_\a d) - \nabla_\b (b^\b_\g c^\g_\a) - \frac{1}{2}\nabla_\b (c^\b_\g b^\g_\a)
+ \frac{3}{2}\nabla_\a (b^\b_\g c^\g_\b) \nonumber \\
&+&  \frac{1}{2}\nabla_\b(b^\b_\g b^\g_\d b^\d_\a) -\frac{1}{2}\nabla_\a(b^\b_\g b^\g_\d b^\d_\b)]t^2,
\label{eq:R0amixed_rad}
\eq
\bq
R^\b_\a &=& -\frac{1}{4t^2}\d^\b_\a - \frac{1}{4t}(4\widetilde{P}^\b_\a + 3b^\b_\a +b\d^\b_\a)
+ (-\frac{5}{2}c^\b_\a + \frac{5}{4}b^\b_\g b^\g_\a - \frac{1}{4}bb^\b_\a - \frac{1}{2}c\d^\b_\a\nonumber \\
&+& \frac{1}{4}b^\d_\g b^\g_\d \d^\b_\a - \k^\b_\a)
+(-\frac{21}{4}d^\b_\a + \frac{7}{2}b^\b_\g c^\g_\a + \frac{7}{4}c^\b_\g b^\g_\a - \frac{7}{4}b^\b_\g b^\g_\d b^\d_\a
- \frac{1}{2}bc^\b_\a + \frac{1}{4}bb^\b_\g b^\g_\a - \frac{1}{2}cb^\b_\a\nonumber \\
&+& \frac{1}{4}b^\d_\g b^\g_\d b^\b_\a - \frac{3}{4}d\d^\b_\a
+ \frac{3}{4}b^\d_\g c^\g_\d \d^\b_\a - \frac{1}{4}b^\d_\g b^\g_\ep b^\ep_\d \d^\b_\a - \l^\b_\a)t,
\label{eq:Rabmixed_rad}
\eq
For the  scalar curvature we find,
\be
R = \dfrac{1}{t}R^{(-1)} + R^{(0)}  + R^{(1)} t,
\label{eq:Rshortsin}
\ee
explicitly,
\bq
R &=& -\frac{1}{t}(\widetilde{P} + 2b) + (-6c -\frac{1}{4}b^2 + \frac{11}{4}b^\b_\a b^\a_\b - \k) \nonumber \\
&+& (-12d -bc + 11b^\b_\a c^\a_\b - \frac{7}{2}b^\b_\g b^\g_\a b^\a_\b + \frac{1}{2}bb^\b_\a b^\a_\b - \l)t.
\label{Rscalar_rad}
\eq
\section{Constraints and evolution}
This Section, which is the heart of this paper,  we work out the asymptotic structure of the various terms appearing in  the system (\ref{eq:Loo}), (\ref{eq:Loa})  (\ref{eq:Lab}) to prepare them for the final counting and balancing done later. In the next Subsection, we express them in terms of the metric coefficients, while later in this Section we arrive at the final asymptotic forms of the splitted equations.

\subsection{Formal expansions}
It is now straightforward to use the previous results to write down analogously  the evolution and constraints in  a splitted form in terms of $a_{\a\b},b_{\a\b},c_{\a\b},d_{\a\b}$. We have:
The hamiltonian constraint (\ref{eq:hamiltonian}) is,
\bq
\mathcal{C}_0 &=& \dfrac{1}{t^3}\left( \dfrac{3}{2}\ep(\widetilde{P} + 2b) \right) \nonumber \\
&+& \dfrac{1}{t^2} \left \{ \dfrac{3}{4} + \ep \left[-\dfrac{1}{2}(\widetilde{P}^2 - 4b^2)
+ \dfrac{3}{2}(-6c - \dfrac{1}{4}b^2 + \dfrac{11}{4}b^\b_\a b^\a_\b - \k)
 \right. \right. \nonumber \\
&+& \left. \left.  2a^{\a\b}\pa_\a \left(\pa_\b (\widetilde{P} + 2b)\right) -2a^{\a\b}\widetilde{\G}^\m_{\a\b}\pa_\m(\widetilde{P} + 2b) \right] \right \} \nonumber \\
&+& \dfrac{1}{t} \left \{ \dfrac{1}{2}(\widetilde{P}+b) + \ep \left[ (\widetilde{P} + 2b)(\dfrac{1}{4}b^\b_\a b^\a_\b - \dfrac{1}{4}b^2 - \k)
- b(-6c -\frac{1}{4}b^2 + \frac{11}{4}b^\b_\a b^\a_\b - \k) \right. \right. \nonumber \\
&+& \left. \left.\dfrac{9}{2}(-12d -bc + 11b^\b_\a c^\a_\b - \frac{7}{2}b^\b_\g b^\g_\a b^\a_\b + \frac{1}{2}bb^\b_\a b^\a_\b - \l)
- 2b^{\a\b}\pa_\a \left(\pa_\b(\widetilde{P}+2b)\right) \right. \right. \nonumber \\
&-& \left. \left. 2a^{\a\b}\pa_\a \left(\pa_\b (-6c -\frac{1}{4}b^2 + \frac{11}{4}b^\d_\g b^\g_\d - \k)\right)
+ 2b^{\a\b}\widetilde{\G}^\m_{\a\b}\pa_\m(\widetilde{P} + 2b) \right. \right. \nonumber \\
&+& \left. \left. 2a^{\a\b}\widetilde{\G}^\m_{\a\b}\pa_\m (-6c -\frac{1}{4}b^2 + \frac{11}{4}b^\d_\g b^\g_\d - \k)
- 2a^{\a\b}E^\m_{\a\b}\pa_\m (\widetilde{P} + 2b) \right] \right \} \nonumber \\
&=& 8\pi G T^0_0.
\label{C0rad all}
\eq
From Eq. (\ref{eq:momentum}), the momentum constraints become,
\bq
\mathcal{C}_\a &=& \dfrac{1}{t^2}\left(-3\ep \pa_\a (\widetilde{P} + 2b)\right) \nonumber \\
&+& \dfrac{1}{t} \left \{ \ep \left[-(\widetilde{P} + 2b)(\nabla_\b b^\b_\a - \nabla_\a b)
+ \pa_\a (-6c -\frac{1}{4}b^2 + \frac{11}{4}b^\d_\g b^\g_\d - \k) \right. \right. \nonumber \\
&-& \left. \left. b^\b_\a \pa_\b (\widetilde{P} + 2b) \right] \right \} \nonumber \\
&+& t \left \{ \frac{1}{2}(\nabla_\b b^\b_\a -\nabla_\a b) + \ep \left[-2(\widetilde{P} + 2b)\big[(\nabla_\b c^\b_\a -\nabla_\a c)
-\frac{1}{2}\nabla_\b (b^\b_\g b^\g_\a) +\frac{1}{2}\nabla_\a (b^\b_\g b^\g_\b)\big] \right. \right. \nonumber \\
&+& \left. \left. (-6c -\frac{1}{4}b^2 + \frac{11}{4}b^\d_\g b^\g_\d - \k)(\nabla_\b b^\b_\a -\nabla_\a b) \right. \right. \nonumber \\
&-& \left. \left. \pa_\a (-12d -bc + 11b^\d_\g c^\g_\d - \frac{7}{2}b^\b_\g b^\g_\d b^\d_\b + \frac{1}{2}bb^\d_\g b^\g_\d - \l)
\right. \right. \nonumber \\
&+& \left. \left. b^\b_\a \pa_\b (-6c -\frac{1}{4}b^2 + \frac{11}{4}b^\d_\g b^\g_\d - \k)
- (2c^\b_\a - b^\b_\g b^\g_\a)\pa_\b (\widetilde{P} + 2b) \right] \right \} \nonumber \\
&=& 8\pi G T^0_\a.
\label{Ca rad all}
\eq
Finally, the snap equation (\ref{eq:pW}) gives,
\bq
&& \dfrac{1}{t^3}\left(-\dfrac{9}{2}\ep (\widetilde{P} + 2b)\right) \nonumber \\
&+& \dfrac{1}{t^2} \left \{ -\dfrac{3}{4} + \ep \left[(\widetilde{P} + 2b)(\dfrac{1}{2}\widetilde{P} + 2b)
-\dfrac{3}{2} (-6c -\frac{1}{4}b^2 + \frac{11}{4}b^\b_\a b^\a_\b - \k) \right. \right. \nonumber \\
&-& \left. \left. 2a^{\a\b}\pa_\a \left(\pa_\b(\widetilde{P} + 2b)\right) + 2a^{\a\b}\widetilde{\G}^\m_{\a\b}\pa_\m(\widetilde{P} + 2b)
+6 a^{\a\b}\pa_\a \left(\pa_\b(\widetilde{P} + 2b)\right) \right. \right. \nonumber \\
&-& \left. \left. 6 a^{\a\b}\widetilde{\G}^\m_{\a\b}\pa_\m(\widetilde{P} + 2b) \right] \right \} \nonumber \\
&+& \dfrac{1}{t} \left \{ -\dfrac{1}{2}\widetilde{P} + \dfrac{3}{2}b
+ \ep \left[\dfrac{9}{2}(-12d - bc + 11b^\b_\a c^\a_\b - \dfrac{7}{2}b^\g_\b b^\b_\a b^\a_\g
+ \dfrac{1}{2}bb^\b_\a b^\a_\b - \l) \right. \right. \nonumber \\
&+& \left. \left. (\widetilde{P} + 2b)(-6c + \dfrac{9}{4}b^\b_\a b^\a_\b - \dfrac{1}{4}b^2 - \k) \right. \right. \nonumber \\
&-& \left. \left. (2\widetilde{P} + 3b)(-6c - \dfrac{1}{4}b^2 + \dfrac{11}{4}b^\b_\a b^\a_\b - \k)
- 6 a^{\a\b}E^\m_{\a\b}\pa_\m (\widetilde{P} + 2b) \right. \right. \nonumber \\
&+& \left. \left. 2b^{\a\b}\pa_\a \left(\pa_\b(\widetilde{P} + 2b) \right)
- 2a^{\a\b}\widetilde{\G}^\m_{\a\b}\pa_\m (-6c - \dfrac{1}{4}b^2 + \dfrac{11}{4}b^\d_\g b^\g_\d - \k) \right. \right. \nonumber \\
&-& \left. \left. 2b^{\a\b}\widetilde{\G}^\m_{\a\b}\pa_\m (\widetilde{P} + 2b)
+ 2a^{\a\b}E^\m_{\a\b}\pa_\m (\widetilde{P} + 2b) \right. \right. \nonumber \\
&-& \left. \left. 6 a^{\a\b}\pa_\a \left(\pa_\b (-6c - \dfrac{1}{4}b^2 + \dfrac{11}{4}b^\d_\g b^\g_\d - \k) \right) \right. \right. \nonumber \\
&-& \left. \left. 6b^{\a\b}\pa_\a \left(\pa_\b (\widetilde{P} + 2b) \right)
+ 6 a^{\a\b}\widetilde{\G}^\m_{\a\b}\pa_\m (-6c - \dfrac{1}{4}b^2 + \dfrac{11}{4}b^\d_\g b^\g_\d - \k) \right. \right. \nonumber \\
&+& \left. \left. 6 b^{\a\b}\widetilde{\G}^\m_{\a\b}\pa_\m (\widetilde{P} + 2b)
+ 2a^{\a\b}\pa_\a \left(\pa_\b(-6c - \dfrac{1}{4}b^2 + \dfrac{11}{4}b^\d_\g b^\g_\d - \k) \right) \right] \right \} \nonumber \\
&=& 8\pi G T^\a_\a,
\label{snap rad all}
\eq
where the components of the energy-momentum tensor $T^i_j$ and its trace are given by the relations (\ref{eq:general perfect fluid}), namely,
\be
T^0_0=\dfrac{1}{3}\r (4{u_0}^2 -1),
\label{eq:Toomixed}
\ee
\be
T^0_\a=\dfrac{4}{3}\r {u_\a}u_0,
\label{eq:Toamixed}
\ee
\be
T^\b_\a=\dfrac{1}{3}\r (4 u^\b u_\a -\d^\b_\a),
\label{eq:Tabmixed}
\ee
and,
\be
T=\textrm{tr}T_{ij}=0.
\label{eq:TrT}
\ee
In view of the identity (\ref{eq:velocities identity}) and the form (\ref{spatial metric rad}), we further get,
\be
1 = u_i u^i \approx {u_0}^2 - \dfrac{1}{t}a^{\a\b} u_\a u_\b .
\label{velocities identity singular}
\ee
Hence,  the constraint and evolution  equations are now in terms of the initial data $a_{\a\b},b_{\a\b},c_{\a\b},d_{\a\b}$. This will allow us later to give a count of how many of the initial data are free.

\subsection{The intermediate asymptotic system}
We now rearrange the terms in the equations given in the previous subsection so as to appear in a form amenable to further simplification. We have:
\be
8\pi G\rho = \mathcal{C}_0 = (L^0_0)^{(-3)}\frac{1}{t^3} + (L^0_0)^{(-2)}\frac{1}{t^2} + (L^0_0)^{(-1)}\frac{1}{t},
\label{eq:C0_s}
\ee
\be
\dfrac{32\pi G}{3}\rho u_\a = \mathcal{C}_\a = (L^0_\a)^{(-2)}\frac{1}{t^2} + (L^0_\a)^{(-1)}\frac{1}{t} + (L^0_\a)^{(0)},
\label{eq:Ca_s}
\ee
and,
\be
-\dfrac{8\pi G}{3}\rho \d^\b_\a = L^\b_\a = (L^\b_\a)^{(-3)}\frac{1}{t^3} + (L^\b_\a)^{(-2)}\frac{1}{t^2} + (L^\b_\a)^{(-1)}\frac{1}{t}.
\label{eq:LAB_s}
\ee
In terms of $a_{\a\b},b_{\a\b},c_{\a\b},d_{\a\b}$,  the quantities appearing in Eq. (\ref{eq:C0_s}) become,
\bq
(L^0_0)^{(-3)} &=& \ep \big[2R^{(-1)}(R^0_0)^{(-2)} - K^{(-1)}R^{(-1)}\big] \nonumber \\
               &=& \dfrac{3}{2}\ep(\widetilde{P} + 2b),
\label{L^0_0(-3)_rad}
\eq
\bq
(L^0_0)^{(-2)} &=& (R^0_0)^{(-2)} + \ep \left[2\left(R^{(-1)}(R^0_0)^{(-1)} + R^{(0)}(R^0_0)^{(-2)}\right) - \dfrac{1}{2}(R^{(-1)})^2  \right. \nonumber \\
&-& \left.  2(\g^{\a\b})^{(-1)}\pa_\a(\pa_\b R^{(-1)}) +2(\g^{\a\b})^{(-1)}(\G^\m_{\a\b})^{(0)}\pa_\m R^{(-1)} - K^{(0)}R^{(-1)}\right] \nonumber \\
&=& \dfrac{3}{4} + \ep \left[-\dfrac{1}{2}(\widetilde{P}^2 - 4b^2)
+ \dfrac{3}{2}(-6c - \dfrac{1}{4}b^2 + \dfrac{11}{4}b^\b_\a b^\a_\b - \k)\right. \nonumber \\
&+& 2a^{\a\b}\pa_\a \big(\pa_\b (\widetilde{P} + 2b)\big)
- \left. 2a^{\a\b}\widetilde{\G}^\m_{\a\b}\pa_\m(\widetilde{P} + 2b) \right],
\label{L^0_0(-2)_rad}
\eq
\bq
(L^0_0)^{(-1)} &=& (R^0_0)^{(-1)} - \dfrac{1}{2}R^{(-1)} + \ep \left \{ 2\left[R^{(-1)}(R^0_0)^{(0)} + R^{(0)}(R^0_0)^{(-1)}
+ R^{(1)}(R^0_0)^{(-2)}\right]  \right. \nonumber \\
&-& \left. R^{(-1)}R^{(0)} - 2\left[(\g^{\a\b})^{(-1)}\pa_\a(\pa_\b R^{(0)}) + (\g^{\a\b})^{(0)}\pa_\a(\pa_\b R^{(-1)})\right] \right. \nonumber \\
&+& \left. 2\left[(\g^{\a\b})^{(-1)}(\G^\m_{\a\b})^{(0)}\pa_\m R^{(0)} + (\g^{\a\b})^{(0)}(\G^\m_{\a\b})^{(0)}\pa_\m R^{(-1)}
 \right. \right. \nonumber \\
&+&\left. \left.  (\g^{\a\b})^{(-1)}(\G^\m_{\a\b})^{(1)}\pa_\m R^{(-1)}\right] +\left[K^{(-1)}R^{(1)} - K^{(1)}R^{(-1)}\right] \right \}  \nonumber \\
&=& \dfrac{1}{2}(\widetilde{P}+b) + \ep \left[ (\widetilde{P} + 2b)(\dfrac{1}{4}b^\b_\a b^\a_\b - \dfrac{1}{4}b^2 - \k)
- b(-6c -\frac{1}{4}b^2 + \frac{11}{4}b^\b_\a b^\a_\b - \k) \right. \nonumber \\
&+& \left. \dfrac{9}{2}(-12d -bc + 11b^\b_\a c^\a_\b - \frac{7}{2}b^\b_\g b^\g_\a b^\a_\b + \frac{1}{2}bb^\b_\a b^\a_\b - \l)
- 2b^{\a\b}\pa_\a \left(\pa_\b(\widetilde{P}+2b)\right) \right. \nonumber \\
&-& \left. 2a^{\a\b}\pa_\a \left(\pa_\b (-6c -\frac{1}{4}b^2 + \frac{11}{4}b^\d_\g b^\g_\d - \k)\right)
+ 2b^{\a\b}\widetilde{\G}^\m_{\a\b}\pa_\m(\widetilde{P} + 2b) \right. \nonumber \\
&+& \left. 2a^{\a\b}\widetilde{\G}^\m_{\a\b}\pa_\m (-6c -\frac{1}{4}b^2 + \frac{11}{4}b^\d_\g b^\g_\d - \k)
- 2a^{\a\b}E^\m_{\a\b}\pa_\m (\widetilde{P} + 2b) \right].
\label{L^0_0(-1)_rad}
\eq
Further, for  Eq. (\ref{eq:Ca_s}), we find
\bq
(L^0_\a)^{(-2)} &=& \ep \big[2\pa_\a R^{(-1)} + (K^\b_\a)^{(-1)}\pa_\b R^{(-1)} \big] \nonumber \\
&=& -3\ep \pa_\a (\widetilde{P} + 2b),
\label{L^0_a(-2)_rad}
\eq
\bq
(L^0_\a)^{(-1)} &=& \ep \left[2R^{(-1)}(R^0_\a)^{(0)} + (K^\b_\a)^{(-1)}\pa_\b R^{(0)} + (K^\b_\a)^{(0)}\pa_\b R^{(-1)} \right] \nonumber \\
&=& \ep \left[-(\widetilde{P} + 2b)(\nabla_\b b^\b_\a - \nabla_\a b)
+ \pa_\a (-6c -\frac{1}{4}b^2 + \frac{11}{4}b^\d_\g b^\g_\d - \k) \right. \nonumber \\
&-& \left. b^\b_\a \pa_\b (\widetilde{P} + 2b) \right],
\label{L^0_a(-1)_rad}
\eq
\bq
(L^0_\a)^{(0)} &=& (R^0_\a)^{(0)} + \ep \left[2\left(R^{(-1)}(R^0_\a)^{(1)} + R^{(0)}(R^0_\a)^{(0)}\right) - 2\pa_\a R^{(1)} \right. \nonumber \\
&+& \left. \left((K^\b_\a)^{(-1)}\pa_\b R^{(1)} + (K^\b_\a)^{(0)}\pa_\b R^{(0)} + (K^\b_\a)^{(1)}\pa_\b R^{(-1)}\right) \right] \nonumber \\
&=& \frac{1}{2}(\nabla_\b b^\b_\a -\nabla_\a b)\nonumber \\ &+& \ep \left[-2(\widetilde{P} + 2b)\big[(\nabla_\b c^\b_\a -\nabla_\a c)
-\frac{1}{2}\nabla_\b (b^\b_\g b^\g_\a) +\frac{1}{2}\nabla_\a (b^\b_\g b^\g_\b)\big] \right. \nonumber \\
&+& \left. (-6c -\frac{1}{4}b^2 + \frac{11}{4}b^\d_\g b^\g_\d - \k)(\nabla_\b b^\b_\a -\nabla_\a b) \right. \nonumber \\
&-& \left. \pa_\a (-12d -bc + 11b^\d_\g c^\g_\d - \frac{7}{2}b^\b_\g b^\g_\d b^\d_\b + \frac{1}{2}bb^\d_\g b^\g_\d - \l)
\right. \nonumber \\
&+& \left. b^\b_\a \pa_\b (-6c -\frac{1}{4}b^2 + \frac{11}{4}b^\d_\g b^\g_\d - \k)
- (2c^\b_\a - b^\b_\g b^\g_\a)\pa_\b (\widetilde{P} + 2b) \right],
\label{L^0_a(0)_rad}
\eq
and finally for  Eq. (\ref{eq:LAB_s}) we have,
\bq
(L^\b_\a)^{(-3)} &=& \ep \big[2R^{(-1)}(R^\b_\a)^{(-2)} + (K^\b_\a)^{(-1)}R^{(-1)} + 4\d^\b_\a R^{(-1)}
- \d^\b_\a K^{(-1)}R^{(-1)} \big] \nonumber \\
&=& -\dfrac{3}{2}\ep \d^\b_\a (\widetilde{P} + 2b),
\label{L^b_a(-3)_rad}
\eq
\bq
(L^\b_\a)^{(-2)} &=& (R^\b_\a)^{(-2)} + \ep \left\{2\left[R^{(-1)}(R^\b_\a)^{(-1)} + R^{(0)}(R^\b_\a)^{(-2)} \right]
-\dfrac{1}{2}\d^\b_\a (R^{(-1)})^2 \right. \nonumber \\
&+& \left. 2(\g^{\b\g})^{(-1)}\pa_\g (\pa_\a R^{(-1)}) - 2(\g^{\b\g})^{(-1)}(\G^\m_{\g\a})^{(0)}\pa_\m R^{(-1)}
+ (K^\b_\a)^{(0)}R^{(-1)} \right. \nonumber \\
&-& \left. 2\d^\b_\a(\g^{\g\d})^{(-1)}\pa_\g (\pa_\d R^{(-1)})
+ 2\d^\b_\a(\g^{\g\d})^{(-1)}(\G^\m_{\g\d})^{(0)} (\pa_\m R^{(-1)}) - \d^\b_\a K^{(0)}R^{(-1)} \right\} \nonumber \\
&=& -\dfrac{1}{4}\d^\b_\a + \ep \left[(\widetilde{P} + 2b)(2\widetilde{P}^\b_\a + \dfrac{1}{2}b^\b_\a + \dfrac{1}{2}b\d^\b_\a
- \dfrac{1}{2}\widetilde{P}\d^\b_\a) \right. \nonumber \\
&-& \left. \dfrac{1}{2}\d^\b_\a (-6c -\frac{1}{4}b^2 + \frac{11}{4}b^\d_\g b^\g_\d - \k) \right. \nonumber \\
&-& \left. 2a^{\b\g}\pa_\g \left(\pa_\a(\widetilde{P} + 2b)\right) + 2a^{\b\g}\widetilde{\G}^\m_{\g\a}\pa_\m(\widetilde{P} + 2b)
+2\d^\b_\a a^{\g\d}\pa_\g \left(\pa_\d(\widetilde{P} + 2b)\right) \right. \nonumber \\
&-& \left. 2\d^\b_\a a^{\g\d}\widetilde{\G}^\m_{\g\d}\pa_\m(\widetilde{P} + 2b) \right],
\label{L^b_a(-2)_rad}
\eq
\bq
(L^\b_\a)^{(-1)} &=& (R^\b_\a)^{(-1)} - \dfrac{1}{2}\d^\b_\a R^{(-1)} + \ep \left\{2\left[R^{(-1)}(R^\b_\a)^{(0)} + R^{(0)}(R^\b_\a)^{(-1)} + R^{(1)}(R^\b_\a)^{(-2)}\right] \right. \nonumber \\
&-& \left. R^{(-1)}R^{(0)}\d^\b_\a + 2\left[(\g^{\b\g})^{(-1)}\pa_\g(\pa_\a R^{(0)}) + (\g^{\b\g})^{(0)}\pa_\g(\pa_\a R^{(-1)}) \right] \right. \nonumber \\
&-& \left. 2\left[(\g^{\b\g})^{(-1)}(\G^\m_{\g\a})^{(0)}\pa_\m R^{(0)} + (\g^{\b\g})^{(0)}(\G^\m_{\g\a})^{(0)}\pa_\m R^{(-1)}
\right. \right. \nonumber \\
&+& \left. \left. (\g^{\b\g})^{(-1)}(\G^\m_{\g\a})^{(1)}\pa_\m R^{(-1)} \right] - (K^\b_\a)^{(-1)}R^{(1)} + (K^\b_\a)^{(1)}R^{(-1)} \right.  \nonumber \\
&-& \left. 2\d^\b_\a \left[(\g^{\g\d})^{(-1)}\pa_\g(\pa_\d R^{(0)})
+ (\g^{\g\d})^{(0)}\pa_\g(\pa_\d R^{(-1)}) \right] \right. \nonumber \\
&+& \left. 2\d^\b_\a \left[(\g^{\g\d})^{(-1)}(\G^\m_{\g\d})^{(0)}\pa_\m R^{(0)}
+ (\g^{\g\d})^{(0)}(\G^\m_{\g\d})^{(0)}\pa_\m R^{(-1)} \right.  \right. \nonumber \\
&+& \left. \left. (\g^{\g\d})^{(-1)}(\G^\m_{\g\d})^{(1)}\pa_\m R^{(-1)} \right] + \d^\b_\a \left(K^{(-1)}R^{(1)} - K^{(1)}R^{(-1)} \right) \right\} \nonumber \\
&=& -\widetilde{P}^\b_\a + \dfrac{1}{2}\widetilde{P}\d^\b_\a - \dfrac{3}{4}b^\b_\a + \dfrac{3}{4}b\d^\b_\a \nonumber \\
&+& \ep \left[\dfrac{3}{2}\d^\b_\a(-12d - bc + 11b^\d_\g c^\g_\d - \dfrac{7}{2}b^\d_\g b^\g_\ep b^\ep_\d
+ \dfrac{1}{2}bb^\d_\g b^\g_\d - \l) \right. \nonumber \\
&+& \left. (\widetilde{P} + 2b)(3c^\b_\a - 3c\d^\b_\a + \dfrac{5}{4}b^\d_\g b^\g_\d \d^\b_\a - \dfrac{1}{4}b^2 \d^\b_\a
- \dfrac{3}{2}b^\b_\g b^\g_\a + \dfrac{1}{2}bb^\b_\a - \d^\b_\a \k + 2\k^\b_\a) \right. \nonumber \\
&-& \left. \dfrac{1}{2}(4\widetilde{P}^\b_\a + 3b^\b_\a + b\d^\b_\a)(-6c - \dfrac{1}{4}b^2 + \dfrac{11}{4}b^\d_\g b^\g_\d - \k)
- 2\d^\b_\a a^{\g\d}E^\m_{\g\d}\pa_\m (\widetilde{P} + 2b) \right. \nonumber \\
&+& \left. 2b^{\b\g}\pa_\g \left(\pa_\a(\widetilde{P} + 2b) \right)
- 2a^{\b\g}\widetilde{\G}^\m_{\g\a}\pa_\m (-6c - \dfrac{1}{4}b^2 + \dfrac{11}{4}b^\d_\ep b^\ep_\d - \k) \right. \nonumber \\
&-& \left. 2b^{\b\g}\widetilde{\G}^\m_{\g\a}\pa_\m (\widetilde{P} + 2b)
+ 2a^{\b\g}E^\m_{\g\a}\pa_\m (\widetilde{P} + 2b) \right. \nonumber \\
&-& \left. 2\d^\b_\a a^{\g\d}\pa_\g \left(\pa_\d (-6c - \dfrac{1}{4}b^2 + \dfrac{11}{4}b^\z_\ep b^\ep_\z - \k) \right) \right.
\nonumber \\
&-& \left. 2\d^\b_\a b^{\g\d}\pa_\g \left(\pa_\d (\widetilde{P} + 2b) \right)
+ 2\d^\b_\a a^{\g\d}\widetilde{\G}^\m_{\g\d}\pa_\m (-6c - \dfrac{1}{4}b^2 + \dfrac{11}{4}b^\z_\ep b^\ep_\z - \k) \right. \nonumber \\
&+& \left. 2\d^\b_\a b^{\g\d}\widetilde{\G}^\m_{\g\d}\pa_\m (\widetilde{P} + 2b)
+ 2a^{\b\g}\pa_\g \left(\pa_\a(-6c - \dfrac{1}{4}b^2 + \dfrac{11}{4}b^\d_\ep b^\ep_\d - \k) \right) \right].
\label{L^b_a(-1)_rad}
\eq
This intermediate form of the constraint and evolution equations is now suitable for further simplification.

\subsection{The asymptotic system}
From Eq. (\ref{eq:C0_s}), the degree of the energy density $\rho$ seems to be equal to $-3$, which is not in accordance with the Friedmann solution (\ref{a prop t^1/2}) and Eq. (\ref{eq ra^4}). However, in this subsection we prove that the term $(L^0_0)^{(-3)}$ is equal to zero, and  the term $(L^0_0)^{(-2)}$ does not vanish. Therefore  the degree of the energy density becomes equal to $-2$. The method of proof below is based on the use of the trace of the higher-order equations to simplify the nine terms $(L^i_j)^{(-n)}$.

In case of radiation ($p=\r /3$), the trace of the field equations (\ref{eq:FEs_rad}) gives the identity,
\be
R - 6\ep \Box_g R = 0.
\label{idBox2}
\ee
Then, for the $t^{-3}$ order terms we get,
\be
\widetilde{P} + 2b = 0,
\label{eq for P}
\ee
for the $t^{-2}$ order terms we obtain,
\be
-12d -bc + 11b^\b_\a c^\a_\b - \frac{7}{2}b^\b_\g b^\g_\a b^\a_\b + \frac{1}{2}bb^\b_\a b^\a_\b - \l = 0,
\label{eq for R1}
\ee
and finally for the $t^{-1}$ order terms we have,
\be
\dfrac{1}{2} a^{\a\b} a^{\m\ep}A_{\a\b\ep}\pa_\m \left( -6c -\frac{1}{4}b^2 + \frac{11}{4}b^\b_\a b^\a_\b - \k \right) = a^{\a\b}\pa_\a \left( \pa_\b (-6c -\frac{1}{4}b^2 + \frac{11}{4}b^\d_\g b^\g_\d - \k)\right).
\label{eq for paR0}
\ee
Using the identities (\ref{eq for P}), (\ref{eq for R1}) and (\ref{eq for paR0}),  the terms $(L^0_0)^{(-3)},(L^0_\a)^{(-2)},(L^\b_\a)^{(-3)}$ vanish identically, while the  six further quantities appearing in (\ref{eq:C0_s})-(\ref{eq:LAB_s}) become,
\be
(L^0_0)^{(-2)} = \dfrac{3}{4} + \dfrac{3}{2}\ep( -6c -\frac{1}{4}b^2 + \frac{11}{4}b^\b_\a b^\a_\b - \k ),
\label{L^0_0(-2)_radnew}
\ee
\be
(L^0_0)^{(-1)} = -\dfrac{1}{2}b -\ep b( -6c -\frac{1}{4}b^2 + \frac{11}{4}b^\b_\a b^\a_\b - \k ),
\label{L^0_0(-1)_radnew}
\ee
and,
\be
(L^0_\a)^{(-1)} = \ep \pa_\a ( -6c -\frac{1}{4}b^2 + \frac{11}{4}b^\d_\g b^\g_\d - \k ),
\label{L^0_a(-1)_radnew}
\ee
\bq
(L^0_\a)^{(0)} &=& \frac{1}{2}(\nabla_\b b^\b_\a -\nabla_\a b)
+ \ep \left[ (-6c -\frac{1}{4}b^2 + \frac{11}{4}b^\d_\g b^\g_\d - \k)(\nabla_\b b^\b_\a -\nabla_\a b) \right. \nonumber \\
&+& \left. b^\b_\a \pa_\b (-6c -\frac{1}{4}b^2 + \frac{11}{4}b^\d_\g b^\g_\d - \k) \right],
\label{L^0_a(0)_radnew}
\eq
and also,
\be
(L^\b_\a)^{(-2)} = -\dfrac{1}{4}\d^\b_\a - \dfrac{1}{2}\ep \d^\b_\a ( -6c -\frac{1}{4}b^2 + \frac{11}{4}b^\d_\g b^\g_\d - \k ),
\label{L^b_a(-2)_radnew}
\ee
\bq
(L^\b_\a)^{(-1)} &=& -\widetilde{P}^\b_\a - \dfrac{3}{4}b^\b_\a - \dfrac{1}{4}b\d^\b_\a \nonumber \\
&+& \ep \left[ \dfrac{1}{2}(4\widetilde{P}^\b_\a + 3b^\b_\a + b\d^\b_\a)(-6c - \dfrac{1}{4}b^2 + \dfrac{11}{4}b^\d_\g b^\g_\d - \k) \right. \nonumber \\
&-& \left. a^{\b\g}a^{\m\ep}A_{\g\a\ep}\pa_\m (-6c - \dfrac{1}{4}b^2 + \dfrac{11}{4}b^\d_\ep b^\ep_\d - \k) \right. \nonumber \\
&+& \left. 2a^{\b\g}\pa_\g \left(\pa_\a(-6c - \dfrac{1}{4}b^2 + \dfrac{11}{4}b^\d_\ep b^\ep_\d - \k)\right) \right].
\label{L^b_a(-1)_radnew}
\eq
These can all be further simplified. Using (\ref{idBox2}) and the identity,
\be
\nabla_0 L^0_j + \nabla_\b L^\b_j=0,
\label{idLij}
\ee
 we find that for $j=\a$,
\be
\pa_t L^0_\a - \dfrac{1}{2}K^\b_\a L^0_\b + \pa_\b L^\b_\a + KL^0_\a - \G^\g_{\a\b}L^\b_\g + \G^\b_{\b\g}L^\g_\a = 0.
\label{identity of L^i_a}
\ee
Substituting their series for each of the various terms  in (\ref{identity of L^i_a}) we find that balancing at the $t^{-2}$ order, the coefficient $(L^0_\a)^{(-1)}$ vanishes identically, namely,
\be
\pa_\a ( -6c -\frac{1}{4}b^2 + \frac{11}{4}b^\d_\g b^\g_\d - \k ) = 0.
\label{id zero paR0}
\ee
Consequently, the terms of (\ref{L^0_a(0)_radnew}) and (\ref{L^b_a(-1)_radnew}) become,
\be
(L^0_\a)^{(0)} = \frac{1}{2}(\nabla_\b b^\b_\a -\nabla_\a b)
+ \ep  (-6c -\frac{1}{4}b^2 + \frac{11}{4}b^\d_\g b^\g_\d - \k)(\nabla_\b b^\b_\a -\nabla_\a b),
\label{L^0_a(0)_radnewfinal}
\ee
and,
\bq
(L^\b_\a)^{(-1)} &=& -\widetilde{P}^\b_\a - \dfrac{3}{4}b^\b_\a - \dfrac{1}{4}b\d^\b_\a \nonumber \\
&-& \dfrac{1}{2} \ep (4\widetilde{P}^\b_\a + 3b^\b_\a + b\d^\b_\a)(-6c - \dfrac{1}{4}b^2 + \dfrac{11}{4}b^\d_\g b^\g_\d - \k).
\label{L^b_a(-1)_radnewfinal}
\eq
Therefore the system of equations (\ref{eq:C0_s})-(\ref{eq:LAB_s}) which describes the dependence between the initial data becomes,
\be
8\pi G\rho = (L^0_0)^{(-2)}\frac{1}{t^2} + (L^0_0)^{(-1)}\frac{1}{t},
\label{eq:C0_sradf2}
\ee
\be
\dfrac{32\pi G}{3}\rho u_\a = (L^0_\a)^{(0)},
\label{eq:Ca_sradf2}
\ee
\be
-\dfrac{8\pi G}{3}\rho \d^\b_\a = (L^\b_\a)^{(-2)}\frac{1}{t^2} + (L^\b_\a)^{(-1)}\frac{1}{t}.
\label{eq:LAB_sradf2}
\ee
Then, from Eq. (\ref{eq:C0_sradf2}) we find one relation for the energy density $\rho$, namely,
\be
8\pi G\rho =  \left [ \dfrac{3}{4} + \dfrac{3}{2}\ep( -6c -\frac{1}{4}b^2 + \frac{11}{4}b^\b_\a b^\a_\b - \k ) \right ] \frac{1}{t^2}
- \left [ \dfrac{1}{2}b +\ep b( -6c -\frac{1}{4}b^2 + \frac{11}{4}b^\b_\a b^\a_\b - \k ) \right ] \frac{1}{t}.
\label{eq for energy density rad with L00(-2) and L00(-1)}
\ee
Substituting $\rho$ from Eq. (\ref{eq for energy density rad with L00(-2) and L00(-1)}) to Eq. (\ref{eq:Ca_sradf2}), we find three more relations for the velocities $u_\a$ and the spatial tensors $a_{\a\b},b_{\a\b},c_{\a\b},d_{\a\b}$, namely,
\be
u_\a = \dfrac{3(L^0_\a)^{(0)}}{4(L^0_0)^{(-2)}}t^2.
\label{final eq of u_a}
\ee
These last relations concerning the velocities $u_\a$ and the spatial tensors $a_{\a\b},b_{\a\b},c_{\a\b},d_{\a\b}$ read,
\bq
u_\a &=& \dfrac{3t^2}{4} \left [ \frac{1}{2}(\nabla_\b b^\b_\a -\nabla_\a b)
+ \ep  (-6c -\frac{1}{4}b^2 + \frac{11}{4}b^\d_\g b^\g_\d - \k)(\nabla_\b b^\b_\a -\nabla_\a b) \right ] \nonumber \\
& \times & \left [ \dfrac{3}{4} + \dfrac{3}{2}\ep( -6c -\frac{1}{4}b^2 + \frac{11}{4}b^\b_\a b^\a_\b - \k ) \right ]^{-1}.
\label{eq for velocities rad with L0a(0) and L00(-2)}
\eq
Continuing, we substitute $\rho$ from Eq. (\ref{eq for energy density rad with L00(-2) and L00(-1)}) to Eq. (\ref{eq:LAB_sradf2}) and we obtain,
\be
(L^\b_\a)^{(-2)}\frac{1}{t^2} + (L^\b_\a)^{(-1)}\frac{1}{t} = -\dfrac{1}{3}\d^\b_\a \left[ (L^0_0)^{(-2)}\frac{1}{t^2} + (L^0_0)^{(-1)}\frac{1}{t} \right].
\label{eq:L00_LAB}
\ee
Using the last equation and the results for the required terms $(L^i_j)^{(-n)}$, we find that the terms of order $t^{-2}$ cancel, while the terms of order $t^{-1}$ give,
\be
-3(\widetilde{P}^\b_\a + \dfrac{3}{4}b^\b_\a + \dfrac{5}{12}b\d^\b_\a)
-\dfrac{1}{2} \ep(12\widetilde{P}^\b_\a + 9b^\b_\a + 5b\d^\b_\a)(-6c - \dfrac{1}{4}b^2 + \dfrac{11}{4}b^\d_\g b^\g_\d
- \k) = 0.
\label{eq:L00_LAB(-1)}
\ee
Notice that the trace of Eq. (\ref{eq:L00_LAB(-1)}) gives the identity (\ref{eq for P}).

\section{The final balance}
In this Section, we arrive at the final number of free functions possible in the radiation solution of the theory without assuming any symmetry. This number (equal to 15 arbitrary functions) is less than that required  (20) for the radiation solution to be a general one in the quadratic theory. Then we show that this number would remain the same had we considered higher-than-forth-order terms in the original metric expansion. This completes the proof of the general structure of the radiation solution in the quadratic  theory.
\subsection{Counting}
From equations (\ref{eq for energy density rad with L00(-2) and L00(-1)}), (\ref{final eq of u_a}) and (\ref{eq:L00_LAB(-1)}), we  already found $1$ relation for the energy density $\rho$, $3$ for the velocities $u^\a$ and additional $6$ concerning the initial data $a_{\a\b},b_{\a\b},c_{\a\b},d_{\a\b}$. Also, we should not make the mistake to count as extra conditions between the initial data the identities (\ref{eq for P})-(\ref{eq for paR0}) as well as the identity (\ref{id zero paR0}), because the trace of the gravitational field equations and the conservation laws yield directly to the same relations.

Nevertheless, we have to take into consideration the fact that the choice of the time $t$ in the metric (\ref{spatial metric rad}) is completely determined by the condition $t=0$ at the singularity, while the space coordinates still permit arbitrary transformations that do not involve the time. These arbitrary transformations can be used, for example, to bring tensor $a_{\a\b}$ to diagonal form.

Thus, from the initial $28$ functions $a_{\a\b},b_{\a\b},c_{\a\b},d_{\a\b},\rho,u^\a$, we have to subtract $13$ that are not arbitrary, and we finally find that the solution contains all together $15$ physically different arbitrary functions (below we also give the possible choice of the initial data that we can use for the particular solution that we find after this analysis). In comparison with the number $20$ which corresponds to a general solution of the problem, we see that after the imposition of the higher-order evolution and constraint equations, the tensor $\g_{\a\b}$ of the form (\ref{spatial metric rad}) cannot correspond to a general solution.

It is also worth noting that setting $\ep=0$, the basic equations
(\ref{eq for energy density rad with L00(-2) and L00(-1)}), (\ref{final eq of u_a}) and (\ref{eq:L00_LAB(-1)}) lead us to conclude that  we  have the exact same results as those found  in general relativity \cite{ll} .

The last question is: Out of the $28$ different functions $a_{\a\b},b_{\a\b},c_{\a\b},d_{\a\b},\rho,u^\a$, which $15$ of those should be chosen as our initial data? We have shown that in the case of radiation the higher-order gravity equations (\ref{eq:pg}), (\ref{eq:pK}), (\ref{eq:pD}) and (\ref{eq:pW}) together with the constraint equations (\ref{eq:hamiltonian}) and (\ref{eq:momentum}) admit a singular formal series expansion of the form  (\ref{spatial metric rad}) leading to a  solution which is not general and requires 15 smooth initial data.

If we prescribe the $28$ data
\be
a_{\a\b}, \quad  b_{\a\b},\quad  c_{\a\b},\quad  d_{\a\b},\quad  \rho,\quad  u^\a
\ee
initially, we still have the freedom to fix $13$ of them. We choose to leave the twelve components of the metrics $c_{\a\b}$ and $d_{\a\b}$ free, and we choose the symmetric space tensor $a_{\a\b}$ to be diagonal. Then we proceed to count the number of free functions in three steps, starting from these $3+0+2\times 6 +1+3=19$ functions. First, (\ref{eq:C0_sradf2}) fixes the function $\rho$ and secondly Eq. (\ref{final eq of u_a}) fixes the $3$ more components of $ u^\a$. Lastly, we use  the 6 relations in (\ref{eq:L00_LAB(-1)}) to completely fix the $6$ components of $b_{\a\b}$. Summing up the free functions we have found, we end up with,
\be
\underbrace{3}_{\textrm{from}\ a_{\a\b}} + \underbrace{0}_{\textrm{from}\ b_{\a\b}}
+\underbrace{12}_{\textrm{from}{\ c_{\a\b}}\ \textrm{and}\ d_{\a\b}} + \underbrace{0}_{\textrm{from}\ \rho}
+ \underbrace{0}_{\textrm{from}\ u^\a} = 15,
\label{eq:notation_initial_data_rad}
\ee
which are suitable free data as required for the specific particular solution.
Obviously this is not the only way to choose among the initial data. For instance, we can choose the space tensor $a_{\a\b}$ not to be diagonal and fix three components of $c_{\a\b}$. Then, we end up with,
\be
\underbrace{6}_{\textrm{from}\ a_{\a\b}} + \underbrace{0}_{\textrm{from}\ b_{\a\b}}
+\underbrace{3}_{\textrm{from}\ c_{\a\b}} + \underbrace{6}_{\textrm{from}\ d_{\a\b}} + \underbrace{0}_{\textrm{from}\ \rho}
+ \underbrace{0}_{\textrm{from}\ u^\a} = 15,
\label{eq:notation_initial_data_rad_2}
\ee
suitable free data. Other choices are also possible.

\subsection{The minimal-order expansion}\label{gthan4}
In this subsection, we show that the number of degrees of freedom of the gravitational field equations in $R + \ep R^2$ theory plus radiation would remain the same if instead of the singular form (\ref{spatial metric rad}) we take a singular expression of the form,
\be
\g_{\a\b}=\sum{\g^{(n)}_{\a\b} t^{n}},\quad n> 4.
\label{gengab2}
\ee
We assume a formal series representation of the spatial metric of the form:
\be
\g_{\a\b} = (\g_{\a\b})^{(1)}t + (\g_{\a\b})^{(2)}t^2 + (\g_{\a\b})^{(3)}t^3 + (\g_{\a\b})^{(4)}t^4 + \cdots
+ (\g_{\a\b})^{(n)}t^n,
\label{general spatial metric rad}
\ee
where $n$ is a natural number, with $n>4$. Obviously, we have,
\be
(\g_{\a\b})^{(1)}=a_{\a\b}, \quad (\g_{\a\b})^{(2)}=b_{\a\b}, \quad (\g_{\a\b})^{(3)}=c_{\a\b}, \quad (\g_{\a\b})^{(4)}=d_{\a\b}.
\label{terms of general spatial metric rad}
\ee
We note that the expression (\ref{general spatial metric rad}) contains $6n$ degrees of freedom ($6$ of each one of the $n$ spatial matrices). Adding $4$ additional degrees of freedom of the energy density $\rho$ and the velocities $u^\a$, the main question now becomes: how many of these $6n+4$ data are independent? (That is when we assume the validity of our dynamical system of equations,  and the metric (\ref{general spatial metric rad}) is taken to be a solution of it, that is of the evolution equations (\ref{eq:pg}), (\ref{eq:pK}), (\ref{eq:pD}) and (\ref{eq:pW}), together with the constraint equations (\ref{eq:hamiltonian}), (\ref{eq:momentum}), and also the radiation equation of state, and relation (\ref{eq:velocities identity}) on each slice $\mathcal{M}_{t}$.)

To answer this question, we consider again our basic system of equations (\ref{eq:C0_sradf2}), (\ref{eq:Ca_sradf2}) and (\ref{eq:LAB_sradf2}) which now due to (\ref{general spatial metric rad}) takes the form,
\be
8\pi G\rho = (L^0_0)^{(-2)}\frac{1}{t^2} + (L^0_0)^{(-1)}\frac{1}{t} + \cdots + (L^0_0)^{(n-5)}t^{n-5},
\label{eq:C0_rad_new2_gen2}
\ee
\be
\dfrac{32\pi G}{3}\rho u_\a = (L^0_\a)^{(0)} + \cdots + (L^0_\a)^{(n-4)}t^{n-4},
\label{eq:Ca_rad_new2_gen2}
\ee
\be
-\dfrac{8\pi G}{3}\rho \d^\b_\a = (L^\b_\a)^{(-2)}\frac{1}{t^2} + (L^\b_\a)^{(-1)}\frac{1}{t} + \cdots
+ (L^\b_\a)^{(n-5)}t^{n-5}.
\label{eq:LAB_rad_new2_gen2}
\ee
Then (\ref{eq:C0_rad_new2_gen2}) gives us $1$ relation for the energy density $\rho$, which when substituted in (\ref{eq:Ca_rad_new2_gen2}) gives $3$ relations for the velocities,
\be
u_\a = \dfrac{3(L^0_\a)^{(0)}}{4(L^0_0)^{(-2)}}t^2.
\label{final eq of u_a_gen}
\ee
This is exactly the same relation between the velocities $u_\a$ and the spatial matrices  as that we found in the case of only using the first four terms of $\g_{\a\b}$. FInally, substituting $\rho$ from (\ref{eq:C0_rad_new2_gen2}) in Eq. (\ref{eq:LAB_rad_new2_gen2}),  we obtain,
\bq
(L^\b_\a)^{(-2)}\frac{1}{t^2} + (L^\b_\a)^{(-1)}\frac{1}{t} + \cdots
+ (L^\b_\a)^{(n-5)}t^{n-5} &=& -\dfrac{1}{3}\d^\b_\a (L^0_0)^{(-2)}\frac{1}{t^2}
-\dfrac{1}{3}\d^\b_\a (L^0_0)^{(-1)}\frac{1}{t} - \cdots \nonumber \\
&-& \dfrac{1}{3}\d^\b_\a(L^0_0)^{(n-5)}t^{n-5}.
\label{eq:L00_LAB_gen}
\eq
The terms of order $t^{-2}$ in (\ref{eq:L00_LAB_gen}) cancel as before, and the $t^{-1}$ order terms give $6$ relations between the initial data, which is the equation (\ref{eq:L00_LAB(-1)}). For the $t^k$-order terms of (\ref{eq:L00_LAB_gen}), where $k=0,1,...,n-5$, we obtain the following equations,
\be
(L^\b_\a)^{(k)} = -\dfrac{1}{3}\d^\b_\a (L^0_0)^{(k)}.
\label{eq:L00_LAB_gen_0,1,...,n-5}
\ee
Apparently, each one of the previous $n-4$ equations gives $6$ relations between  the initial data. Therefore, from (\ref{eq:L00_LAB_gen_0,1,...,n-5}) we find $6\times(n-4)=6n-24$ relations between the data $(\g_{\a\b})^{(0)},...,(\g_{\a\b})^{(n)}$ and so, counting the $t^{-2}$ and $t^{-1}$ order terms together with the next $n-4$ terms from (\ref{eq:L00_LAB_gen}) we get $0+6+(6n-24)=6n-18$ relations in total.
Consequently, taking further into account one relation from (\ref{eq:C0_rad_new2_gen2}), three relations from (\ref{final eq of u_a_gen}) and $(6n-18)$ relations from (\ref{eq:L00_LAB_gen}), the counting give us $1+3+(6n-18)=6n-14$ relations between the initial data. Subtracting this from the total $6n+4$ data that we started from, we conclude that only $18$ functions can be arbitrary. Hence, subtracting $3$ diffeomorphisms gives the final number 15 of  free functions for the problem. Accordingly, we have  shown that without loss of generality we can stop the series  (\ref{general spatial metric rad}) at the order $4$.

\section{Discussion}
In this paper we have analysed the genericity aspect of radiation cosmologies in quadratic gravity. We have shown that perturbations of the general relativistic radiation isotropic solution in the framework of the quadratic theory cannot sustain a generic radiation solution because the final solution has fewer than necessary arbitrary functions than those which must be present in a general radiation solution in quadratic gravity. The unperturbed GR radiation isotropic solution ($a\sim \sqrt{t}$) as a solution in quadratic gravity is consequently non-generic and unstable, and so we expect it to be replaced by another solution. Starting with a radiation stage at early times as in the standard cosmological model, our results indicate that there must be a future replacement, and the radiation-matter transition in the standard model in the quadratic theory might be justified in this light. Similarly, a transition from a radiation regime to an earlier one such as an inflationary stage, is also dependent on the instability of the former as in the present work.

There are two aspects of these results worth noting. The first is that although not general, our solution contains 15 arbitrary functions and so we expect  slightly inhomogeneous radiation solutions to have a similar instability and decay into something else (either in the future or past direction). The second feature  is that the non-generality of the radiation solution will hold true for any polynomial lagrangian theory, not just the quadratic, for instance a theory like $R+\ep R^n,n>2$. This follows from an examination of the various terms and we may conclude that $n$-th order terms will not change the qualitative nature of the final result.

There is finally the question of what is the nature of the general radiation solution with 20 arbitrary functions (instead of the one with 15 data constructed  here) in the framework of the quadratic theory? Such a generic radiation solution is predicted by the analysis of this paper, but its construction requires techniques beyond those advanced here. A radiation solution with 20 arbitrary functions would necessarily be global and may not be constructed perturbatively, as the present work implied. It is unknown whether such generic radiation models will necessarily be singular or not.

\section*{Acknowledgments}
We thank S.~D. Odintsov for useful  correspondence.
This paper is dedicated to the fond memory of our beloved friend and colleague, the late George Flessas. George was a very kind and vibrant man who always  encouraged young people to pursue their scientific  dreams. He had many interests in mathematical physics, and perhaps above all he was a true master in handling special functions and differential equations. He played an instrumental role in the statement of the past-instability conjecture.

\end{document}